\documentclass[aps,prb, superscriptaddress,twocolumn,showpacs]{revtex4}

\usepackage{graphicx}
\usepackage{rotating}
\usepackage{longtable} 
\usepackage[usenames]{color}

\begin{document} 

\newcommand{\lsmo}{\ensuremath{\mathrm{La_{0.5}Sr_{1.5}MnO_4}}}
\newcommand{\oo}{($\frac{1}{4}$,$\frac{1}{4}$,0)}
\newcommand{\mo}{($\frac{1}{4}$,$-\frac{1}{4}$,$\frac{1}{2}$)~}
\newcommand{\lsmobi}{\ensuremath{\mathrm{LaSr_{2}Mn_{2}O_7}}~}

\title{Surface magnetism of Rh(001) from LDA+U calculations }

\author{N. Stoji\' c}
\affiliation{   Abdus Salam International Centre for Theoretical Physics,
                         Trieste 34014,
                         Italy}
\author{N. Binggeli}
\affiliation{   Abdus Salam International Centre for Theoretical Physics,
                         Trieste 34014,
                         Italy}
\affiliation{  INFM Democritos National Simulation Center,
                       Trieste I-34014, Italy }
\author{M. Altarelli}
\affiliation{   Abdus Salam International Centre for Theoretical Physics,
                         Trieste 34014,
                         Italy}
\affiliation{  European XFEL Project Team,
                       Desy,
                       Notkerstra\ss e 85,
                       22607 Hamburg, Germany   }

\date{\today}

\begin{abstract}

We report calculations indicating the presence of a surface magnetic moment for Rh(001), motivated
by the detection of a finite moment by  magnetic linear dichroism experiments.
 We show that, while the density
functional with the local density or generalized gradient approximations (LDA and GGA)
for exchange and correlation
yields a non-magnetic ground state, the application of the GGA plus on-site Coulomb interaction $U$
method predicts surface magnetism, thus offering
 a solution to the long-standing discrepancy between experiment and theory. The calculated 
moment on the outermost Rh atom increases with the strength of the effective on-site parameter
$U_{\rm eff}=U-J$, for  $U_{\rm eff}\geq 1.2$~eV, and is as large as 1.24~$\mu_B$ for $U_{\rm eff}=2.5$~eV.

\end{abstract}


\maketitle

The finding of a surface moment for a  material which is non-magnetic in the bulk could have far-reaching
consequences. It would give valuable directions for design of novel nanostructured materials
and  could have technological implications for the magnetic storage industry in the future.
 So far, the only experimental evidence of a magnetic surface on a non-magnetic material
has been obtained for the rhodium (001) surface. \cite{GolBarCom99} Goldoni {\it et al.}\cite{GolBarCom99}
measured the linear magnetic dichroism in $3d$ core-level photoemission  of  Rh(001) surface.
They applied an external magnetic field and reversed it by 180$^\circ$, which 
caused a difference in the core level intensity and shape, indicating the presence of 
a magnetic moment. In a later study,\cite{GolBarBar00} the same authors 
concluded that the rhodium (001) surface displays either a weak ferromagnetic
ordering or super-paramagnetism at 100~K. The conclusion is based on their finding that
a small residual magnetic field is necessary for the dichroic signal, but, at the same time, the
intensity of the effect does not depend on the magnitude of the applied residual field.
Earlier spin-polarized photoemission study by Wu {\it et al.}\cite{WuGarBeg94} 
found a weakly ferromagnetic Rh(001) surface at room temperature with an estimated moment per surface atom between
0.1~$\mu_B$ and 0.2~$\mu_B$.

The experimental findings are in line with the expectation that the 
tendency towards magnetism increases near metal surfaces, because of the narrowing of the density of states
which yields a Stoner enhancement in the susceptibility. The elements which are close to satisfy 
Stoner criterion in the bulk phase are likely to form magnetic ordering in a reduced dimensionality. 
In particular, there are theoretical and experimental efforts directed towards investigation of
surface magnetism in some $4d$ and $5d$  transition metals. For a few decades now, many theoretical studies
have examined the properties of rhodium surfaces\cite{EicHafFur96} and possibilities of 
surface magnetism in rhodium. \cite{ChoSch97,ChoKan95,Cip00,BarGuiSpa00}
Rhodium clusters have been investigated theoretically\cite{RedKhaDun93,WilSteLan95,AguRodMic02}
and experimentally.\cite{CoxLouAps94} Overall, it has been found that the moment depends on the size and the
geometrical structure of the cluster, with the calculated values of the moments somewhat larger
than the measured value of 0.8~$\mu_B$ per atom for a ten-atom cluster.\cite{CoxLouAps94}
Rh monatomic wires have also been shown theoretically to be magnetic.\cite{DelTos04}
Calculations of rhodium overlayers on 
Ag and Au\cite{EriAlbBor91,Blu92} yielded a moment on Rh atoms. 
Recently, ferromagnetic order has been found theoretically in bulk Rh at the
optimum lattice constant in the body centered cubic lattice.\cite{HugOsu04}
So far, no realistic {\it ab-initio} study has found a
ferromagnetic solution for the rhodium (001) surface. Cho and Scheffler \cite{ChoSch97}
found a non-magnetic ground state which  was practically degenerate in energy with the ferromagnetic state,
obtained as a solution of the
constrained, fixed spin-moment calculations, for moments below 0.5~$\mu_B$.
Similar results have been obtained more recently,\cite{Cip00} with an imposed moment limited
to the surface layers. These calculations indicate a high susceptibility on the rhodium surface and suggest that
surface structures with defects and reconstructions might be magnetic.
Cho and Scheffler \cite{ChoSch97} also solved a long-standing
problem of anomalously large theoretical surface relaxations compared to the experimental
data, by taking into account the thermal expansion of the surface, within the quasi-harmonic
approximation.

In this paper, we explore
the possibility for magnetism in rhodium (001) surface within the framework of the
density-functional theory (DFT), using the generalized gradient approximation (GGA) of the
exchange-correlation potential, and the GGA plus on-site Coulomb interaction approach (denoted
as LDA+U from this point on).
As the discrepancy between theoretical and experimental results on the surface magnetism
of rhodium persists for all previously attempted band-structure approaches, it seems likely that 
the approximation used for exchange and correlation may be the reason for 
the non-magnetic solution. There are various examples in
the literature, not limited to  strongly correlated systems, for which the LDA+$U$ was originally
designed,  where application of this method corrected the DFT solutions by 
yielding results in agreement with experiments.\cite{MohPerBla01}
The  LDA+$U$ is a method going beyond
the LDA by treating exchange and correlation differently for a chosen set of states, in this case, 
the rhodium $4d$ orbitals. The selected orbitals are treated with an orbital dependent
potential with an associated on-site Coulomb and exchange interactions, $U$ and $J$. In the LDA,
the electron-electron interactions have already been accounted for in a mean field way,
and, therefore, one needs to apply a double counting correction. There are several existing
versions of this correction.\cite{AniZanAnd91,AniSolKor93,CzySaw94}  We apply the most commonly used one, 
 introduced by Anisimov {\it et al.}\cite{AniSolKor93} 
which satisfies the LDA atomic-like limit to the total energy. 
 It is known as the self-interaction corrected (SIC) LDA+$U$ method.

We solve the DFT equations using the WIEN2k implementation \cite{BlaSchMad01} of the full potential linear
augmented plane wave (FLAPW) method in a supercell geometry. We model the surface using 
an 11-layer slab with a vacuum
thickness corresponding to 6 layers. For the surface layer relaxation, we applied the
previously calculated result\cite{ChoSch97} of $-1.4$~\%, compatible with the
experimental finding of $-1.16\pm 1.6$~\% \cite{BegKimJon93}. We used the optimized lattice
constant of 3.84~\AA,  sphere radius of 1.28~\AA, energy cutoff equal to 13.8~Ry and $k-$point sampling
with (22 x 22 x 1) $k-$points mesh in the full Brillouin zone (66 k-points in the reduced
Brillouin zone). 
The calculations have been performed utilizing the GGA in the form given by Perdew, Burke and
Ernzerhof.\cite{PerBurErn96} From convergence tests with the number of  $k-$points and
energy cutoff, we estimate the numerical accuracy of the energy difference between magnetic and non-magnetic
solution to be 2~meV and of the surface magnetic moment to the 0.02~$\mu_B$. We have also assessed the effect of 
the topmost layer relaxation on the energy difference
 and the surface moment. Changing the relaxation to
-2.8~\%, changes the surface magnetic moment by 0.05~$\mu_B$ and the 
 difference between magnetic and non-magnetic surface energy by 6~meV 
per surface atom within the LDA+$U$.

Considering the fact that the Stoner criterion for bulk rhodium is close to being
fulfilled, we first checked for the effects of the inclusion, within the GGA, of the spin-orbit interaction
in Rh(001) surface calculations, which was not included in previous calculations. 
It did not result in a surface moment. A similar conclusion
was obtained by checking the influence of steps and line defects on the surface, within the GGA. The removal of every
other line, or a pair of lines of surface atoms, parallel with the [110]-direction, gave only a non-magnetic solution. 
Structures with a few lines of surface atoms removed, simulating a step, also yielded no magnetic moment.

\renewcommand{\baselinestretch}{1}
\begin{table}[ht]
\bigskip
\begin{center}
\begin{tabular}{ c c c c }
\hline
\hline
  $U_{\rm eff} (eV)$  &  $m_{5} (\mu_B)$ &   $m_{4} (\mu_B)$ & $ \Delta E_{\rm surf}$ (meV / surf.unit cell) 
    \\
\hline
 $0.5$ & $0$ & $0$ & $-$ \\
 $1.0$ & $0$  & $0$ & $-$ \\
 $1.2$ & $0.07$ & $0.01$ & $-1.8$ \\
 $1.5$ & $0.55$ & $0.16$ & $-5.0$ \\
 $2.0$ & $0.86$ & $0.37$ & $-21.0$ \\
 $2.5$ & $1.24$ & $0.89$ & $-57.1$ \\
\hline
\hline
\end{tabular}
\caption{Magnetic moments on the surface ($m_5$) and subsurface ($m_4$) layers, and difference of
surface energies between magnetic and non-magnetic solutions, $\Delta E_{\rm surf}$,  as a function of the effective Coulomb parameter $U_{\rm eff}$. }
\label{Table_moments}
\end{center}
\end{table}

In contrast, the LDA+$U$ method did induce a magnetic moment on the clean, defect-free surface. 
Table~\ref{Table_moments} 
gives the information on how the magnetic moments in the surface and subsurface layers
change as a function of the strength of the on-site parameter
$U_{\rm eff}$, given by: $U_{\rm eff}=U-J$.\cite{SawMorTer97,DudBotSav98} 
We estimated the values of the Coulomb and exchange parameters for bulk rhodium
utilizing the  Tight Binding Linear Muffin-tin orbital 
code\cite{JepKriBur94} in the
atomic sphere approximation (TB-LMTO-ASA) and obtained $U\approx3.4$~eV and $J\approx0.6$~eV, 
i.e. $U_{\rm eff}=2.8$~eV.  Solovyev {\it et al.} \cite{SolDedAni94}
calculated $U$ and $J$ using the same TB-LMTO code for rhodium impurities in Rb. They obtained a value of 
$U=3.6$~eV, $J=0.6$~eV, $U_{\rm eff}=3.0$~eV for the monovalent Rh$^{1+}$ impurity, 
which is not too different from the value we find. 
It is known that the calculated values of $U$ and $J$ tend to depend somewhat on the method used to calculate them
and consequently have relatively large error bars (of the order of 1~eV).
Therefore, the calculated values are to be used
 as a guidance towards a reasonable physical range of $U$ values and an effort should be made
to observe and analyze trends as a function of $U$. 
In Table~\ref{Table_moments}, we reported the magnetic moments calculated for $U_{\rm eff}$ up to 2.5~eV, together 
with the corresponding difference between the magnetic and non-magnetic surface energies:
$ \Delta E_{\rm surf} = E_{\rm surf}^{FM}- E_{\rm surf}^{NM}$. The
non-magnetic energies were obtained using fixed-spin calculations with moment set to zero. 

For values of $U_{\rm eff}$ below 1.2~eV there is no ferromagnetic solution. 
Starting with $U_{\rm eff}$ of 1.2~eV, the ferromagnetic solution has lower energy than the non-magnetic state. 
The surface energy difference for the case with $U_{\rm eff}=2.5$~eV is significantly larger
than thermal-fluctuation energy at room temperature, thus indicating that the effect could be
observable even at higher temperatures. With the increase of $U_{\rm eff}$ up to 2.5~eV, the
magnetic moment on the surface and subsurface layers increases, and the ferromagnetic solution 
becomes more stable.  We note that already for  $U_{\rm eff}\approx1.5$~eV, our calculated 
magnetic moments are one to two orders of magnitude greater than those obtained by Niklasson et al.\cite{NikMirSkr97}
from LDA+$U$ calculations for Rh overlayers of comparable thickness on Ag(001), and do not disappear with increasing
thickness\cite{note4}. 

\begin{figure}[h]
  \begin{center}
  \includegraphics[width=7cm, angle=270]{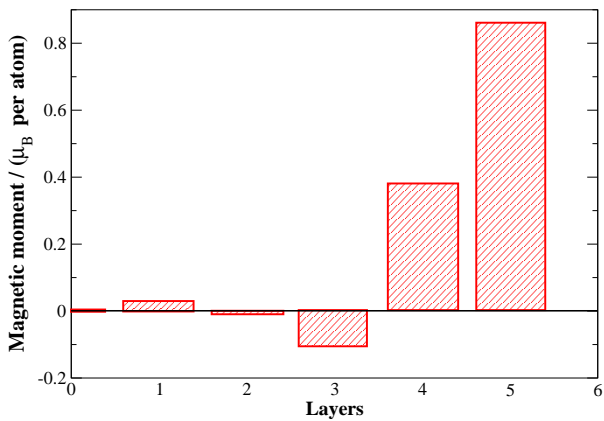}
  \caption{(Color online) Distribution of magnetic moments per atom inside a 11-layer Rh(001) slab. Layer 0 is the center
 and layer 5 the surface of the slab.    }
  \bigskip
  \label{fig:moments}
  \end{center}
\end{figure}

The magnetic moments per atom are 
presented in Fig.~\ref{fig:moments}   as a function of the
layer position inside the slab  for the case  $U_{\rm eff}=2$~eV. The center of slab (denoted as layer 0)
has a vanishing  magnetic moment and the values of the moment are clearly increasing towards the subsurface and
surface layers (layers 4 and 5).
We have also performed  LDA+$U$ calculations
on  bulk rhodium and found that, for all values of $U_{\rm eff}$ used in this work, the non-magnetic solution is
the stable state for the bulk.
Considering the fact that the Rh $4d$ orbitals are not very localized, we performed a test using a
different basis for the LDA + $U$ in a pseudopotential
plane-wave code.\cite{BarDalGir} In this case, the localized-orbital basis consisted of the orthonormal 
atomic tight-binding orbitals, while in the FLAPW approach, the orbitally dependent potential
was applied to the atomic truncated orbitals (and was zero outside the muffin-tin radius).    
In the plane-wave scheme, we used an ultrasoft pseudopotential\cite{RapRabKax90} with a nonlinear core correction 
and Perdew-Burke-Ernzerhof exchange and correlation potential.\cite{PerBurErn96} 
 The calculations gave similar results, with the values of magnetic moments somewhat larger
than in the FLAPW code.  For the sake of completeness,
we did  tests using the LDA for exchange and correlation and keeping all the other parameters
of calculation the same. 
Just like in the case of GGA, the  magnetic moment on the surface atom increases  and the
ferromagnetic solution becomes more stable  with increasing  $U_{\rm eff}$, albeit with slightly smaller
value of the moment on the surface Rh atom.\cite{note2} 
Finally, we also checked for the robustness of the Rh surface magnetism using another version of the LDA+$U$,
with a different double counting correction, implemented in
the WIEN2k code, the so-called mean-field correction.\cite{AniZanAnd91} For somewhat higher values of $U_{\rm eff}$  
it also resulted in a ferromagnetic solution.\cite{note1} Hence, in all cases we obtain a stable surface magnetic
solution, with a significant moment on the outermost atomic layer, for values of  $U_{\rm eff}$ smaller than, or
comparable to the calculated  $U_{\rm eff}$. Larger values of the on-site interaction  $U_{\rm eff}$ induce thicker
magnetic surface regions, requiring the use of thicker slabs.
 
\begin{figure}[h]
  \begin{center}
  \includegraphics[width=7.7cm, angle=0]{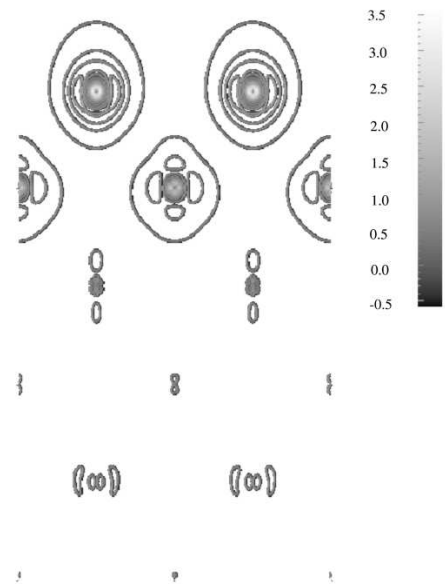}
  \caption{  Spin density difference map (in e/\AA$^3$), for the case  $U_{\rm eff}$=2~eV,
  within a (100) atomic plane perpendicular to the surface.  }
  \bigskip
  \label{fig:spin_map} 
  \end{center}
\end{figure}

To gain a further insight into the microscopic arrangement of magnetic moments, in  Fig.~\ref{fig:spin_map} we show
 the spin density difference map within a (100) atomic plane perpendicular to the surface
for the case $U_{\rm eff}=2$~eV. The spin density is very localized on the atoms of
 the surface and subsurface layers and is
vanishing towards the center of the slab (bottom line of atoms in Fig.~\ref{fig:spin_map}). 
The spin polarized density on the outermost atoms derives mostly from $d_{xy}$,  $d_{3z^2-r^2}$, $d_{xz}$ and $d_{yz}$
orbitals, with a contribution of $\sim$~30\% from the $d_{xy}$ orbitals and of 21-22~\% from each of the other
three types of $d$ orbitals. In fact, these four types of $d$ orbitals dominate the local density of states
on the Rh surface atoms at energies around the Fermi energy (with comparable weights), both when $U_{\rm eff}=0$
and at finite $U_{\rm eff}$ (2.5~eV). Inspection of Fig.~\ref{fig:spin_map} also indicates that the states
which contribute to the spin density map have predominantly antibonding character, which is consistent
with the character of the $d$ states found near the Fermi energy in bulk Rh. 

Intuitively, the magnetic ground state of the LDA+$U$ method in this system
 can be understood by considering the effect of the on-site Coulomb interaction:
it tends to favor the solution in which two spin states of the same  $d$-orbital are separated in energy, 
with an energy separation roughly equal to $U$.
To additionally check the reliability and stability of this method and its implementation, we
applied it to (001) [or (0001) for hexagonal systems] surfaces of some $4d$ and $5d$-elements, namely Mo, Ru,  W and Ir. All of these surfaces
were found to be non-magnetic. We used a relaxed geometry for all systems and, where appropriate,
we took into account the surface reconstruction (Mo and W). In all cases we used the $U-J$ values estimated in 
Ref.~\onlinecite{SolDedAni94} which should be an upper bound for $U_{\rm eff}$.

At present, a truly  quantitative comparison between the experimental results and theory is not
possible. Our solution, for which energy difference is greater than room temperature
fluctuations ($U_{\rm eff}$=2.5~eV), has a magnetic moment on the outermost Rh atom of 1.24~$\mu_B$ at 0~K.
 Goldoni {\it et al.}\cite{GolBarCom99,GolBarBar00} could not give a 
 measure of surface moment,
but concluded that their results are consistent with weakly ferromagnetic or superparamagnetic
surface. The only other experiment\cite{WuGarBeg94}  estimated the surface moment to be between 0.1 and 0.2~$\mu_B$.
Experiments were done at 100~K and room temperature, respectively. As we are unable at present  to 
include the temperature effects in our calculations, numerical comparison is not justified. 
It is also possible that a stronger ferromagnetic ordering in the experiment of 
Goldoni {\it et al.}\cite{GolBarCom99} would be detected, had the $4d$
states been probed directly. Their $3d$ core-level photoemission experiment probed the $4p$ polarization
induced by the $4d$ moment.

In summary, we have shown that a magnetic ground state for the rhodium  (001) surface
is obtained by using the LDA+$U$ approach.
A stable solution with a significant magnetic moment on the outermost Rh atom 
is obtained for $U_{\rm eff}$ smaller than, or comparable to the calculated $U_{\rm eff}$,  
with a trend of increasing
moment and stability of the magnetic solution with increasing $U_{\rm eff}$. 
We believe that our results on the rhodium (001) surface  prove that the 
corrected non-local exchange and correlation
in the LDA+$U$ method is the key ingredient towards an accurate solution in density functional formalism
and in agreement with experiment for this relatively weakly correlated system. It is our hope that similar
approaches could be applied  to other problems
related to magnetism, even in the cases of relatively weakly correlated materials and including systems of lower
dimensions\cite{WieDelTos04}, in situations where the DFT method itself cannot fully reproduce the experimental results. 

\acknowledgments
We would like to thank M. Komelj, G. Paolucci and G. Comelli for helpful and stimulating discussions.
Calculations in this work have been done using, among other codes, the PWscf package. \cite{BarDalGir}
We acknowledge support for this work by the INFM within the framework ``Iniziativa Trasversale Calcolo Parallelo''.

\bibliography{rho}

\begin{thebibliography}{34}
\bibitem[{not({\natexlab{a}})}]{note4}
\bibinfo{note}{We calculated a 13-layer slab and found its surface carries a
  magnetic moment similar to the moment at the surface of an 11-layer slab.}
\bibitem[{not({\natexlab{b}})}]{note2}
\bibinfo{note}{For $U_{\rm eff}=2$~eV, the magnetic moment on the surface Rh
  atom, $m_5$, equals 0.44~$\mu_B$ and the $\Delta E_{\rm surf}$ is $-9.4$~meV.
  For $U_{\rm eff}=2.2$~eV, $m_5=0.62~\mu_B$ and $\Delta E_{\rm
  surf}=-15.2$~meV.}
\bibitem[{not({\natexlab{c}})}]{note1}
\bibinfo{note}{The first magnetic solution was observed for $\sim U_{\rm
  eff}=3$~eV, with $m_5= 0.10~\mu_B$, but with virtually degenerate energies of
  magnetic and non magnetic solutions. For $U_{\rm eff}=3.5$~eV, the moment on
  the outermost Rh atom raised to 0.18~$\mu_B$, and $\Delta E_{\rm
  surf}=-5.4$~meV}.
\end{thebibliography}

\end{document}